\documentclass[superscriptaddress,twocolumn,aps,prl,preprintnumbers,longbibliography]{revtex4-1}
\usepackage{amsfonts,amsmath,amssymb}
\usepackage{graphicx,natbib,bm,hyperref,color}
\usepackage[anythingbreaks]{breakurl}
\usepackage{ulem}
\graphicspath{{./fig/}{./png/}}
\newcommand{\EQ}{\begin{equation}}
\newcommand{\EN}{\end{equation}}
\newcommand{\EQA}{\begin{eqnarray}}
\newcommand{\ENA}{\end{eqnarray}}

\newcommand{\Eq}[1]{Eq.~(\ref{#1})}
\newcommand{\Eqs}[2]{Eqs.~(\ref{#1}) and~(\ref{#2})}

\newcommand{\Fig}[1]{Fig.~\ref{#1}}

\newcommand{\aver}[1]{\langle #1\rangle}

{}
{}
{}

{}
{}
{}
{}
{}
{}
{}
{}
{}
{}
{}
{}
{}
{}
{}
{}
{}

{}

{}

{}
{}
{}

\newcommand{\kk}{\bm{k}}

\newcommand{\xx}{\bm{x}}

\newcommand{\rr}{\bm{r}}

\newcommand{\BB}{\bm{B}}

\newcommand{\JJ}{\bm{J}}

\newcommand{\AAA}{\bm{A}}

\newcommand{\uu}{\bm{u}}

\newcommand{\nab}{{\bm{\nabla}}}

\newcommand{\SSSS}{\mbox{\boldmath ${\sf S}$} {}}

\newcommand{\ii}{{\rm i}}

\newcommand{\DD}{{\rm D} {}}

\newcommand{\dd}{{\rm d} {}}

\newcommand{\const}{{\rm const}  {}}

\def\Sp{\mbox{\rm Sp}}

\def\EEM{{\cal E}_{\rm M}}

\def\HHM{{\cal H}_{\rm M}}

\def\EM{E_{\rm M}}

\def\xiM{\xi_{\rm M}}

\def\HM{H_{\rm M}}
\def\EM{E_{\rm M}}

\def\kB{k_{\rm B}}

\def\kB{k_{\rm B}}

\newcommand{\s}{\,{\rm s}}

\newcommand{\cm}{\,{\rm cm}}

\def\apj{Astrophys. J.}

\def\prd{Phys. Rev. D}

\def\prl{Phys. Rev. Lett.}

\def\mnras{Mon. Not. R. Astron. Soc.}

\def\physrep{Phys. Rep.}

\def\jcap{J. Cosmol. Astropart. Phys.}
\def\jhep{J. High. Energy Phys.}
\def\gafd{Geophys. Astrophys. Fluid Dynam.}
\def\joss{J. Open Source Softw.}
\begin{document}

\date{Received 1 February 2023; accepted 4 April 2023; published 11 May 2023}
\preprint{NORDITA-2023-001, RESCEU-1/23}
\title{Decay law of magnetic turbulence with helicity balanced by chiral fermions}

\author{Axel Brandenburg}
\affiliation{Nordita, KTH Royal Institute of Technology and Stockholm University, 10691 Stockholm, Sweden}
\affiliation{The Oskar Klein Centre, Department of Astronomy, Stockholm University, AlbaNova, 10691 Stockholm, Sweden}
\affiliation{School of Natural Sciences and Medicine, Ilia State University, 0194 Tbilisi, Georgia}
\affiliation{McWilliams Center for Cosmology and Department of Physics, Carnegie Mellon University, Pittsburgh, Pennsylvania 15213, USA}

\author{Kohei Kamada}
\affiliation{Research Center for the Early Universe (RESCEU), Graduate School of Science,\\ The University of Tokyo, Hongo 7-3-1, Bunkyo-ku, Tokyo 113-0033, Japan}\

\author{Jennifer Schober}
\affiliation{Institute of Physics, Laboratory of Astrophysics, \'Ecole Polytechnique F\'ed\'erale de Lausanne (EPFL), 1290 Sauverny, Switzerland}

\begin{abstract}
In plasmas composed of massless electrically charged fermions,
chirality can be interchanged with magnetic helicity while preserving
the total chirality through the quantum chiral anomaly.
The decay of turbulent energy in plasmas such as those in the early Universe
and compact stars is usually controlled by certain conservation laws.
In the case of zero total chirality, when the magnetic helicity density
balances with the appropriately scaled chiral chemical potential to zero,
the total chirality no longer determines the decay.
We propose that in such a case, an adaptation to the Hosking integral,
which is conserved in nonhelical magnetically dominated turbulence,
controls the decay in turbulence with helicity balanced by chiral
fermions.
We show, using a high resolution numerical simulation, that
this is indeed the case.
The magnetic energy density decays and the correlation length increases
with time just like in nonhelical turbulence with vanishing chiral
chemical potential.
But here, the magnetic helicity density is nearly maximum and shows a
scaling with time $t$ proportional to $t^{-2/3}$.
This is unrelated to the $t^{-2/3}$ decay of magnetic {\it energy}
in fully helical magnetic turbulence.
The modulus of the chiral chemical potential decays in the same fashion.
This is much slower than the exponential decay previously expected 
in theories of asymmetric baryon production from the hypermagnetic
helicity decay after axion inflation.
\end{abstract}

\maketitle

Magnetic helicity characterizes the knottedness of magnetic field lines
and plays important roles in cosmological, astrophysical, and laboratory plasmas.
Since the early work of Woltjer of 1958 \cite{Wol58}, we know that
the magnetic helicity is an invariant of the ideal magnetohydrodynamic
(MHD) equations.
Even in the non-ideal case of finite conductivity, it is asymptotically
conserved in the limit of large magnetic Reynolds numbers \citep{Tay74}.
This is because, unlike the magnetic energy dissipation, which is finite
at large magnetic Reynolds numbers, the magnetic helicity dissipation
converges to zero in that limit \citep{BS05}.
The magnetic helicity controls the decay of magnetic fields in closed
or periodic domains, provided the magnetic helicity is finite.
However, even when the net magnetic helicity over the whole volume
vanishes, there can still be random fluctuations of magnetic helicity.
In this case, the conservation of magnetic helicity still plays
an important role, but only in smaller subvolumes, as was shown
recently \citep{Hosking+Schekochihin21}.
The conserved quantity in that case is what is now known as the Hosking
integral \citep{Scheko22, Zhou+22}, which characterizes magnetic helicity
fluctuations in smaller subvolumes \citep{Hosking+Schekochihin21}.

At relativistic energies, the chirality of fermions combines with
the helicity of the magnetic field to a total chirality that is
{\it strictly} conserved in a periodic or closed domain -- even for finite
magnetic diffusivity \citep{BFR12, Roga_etal17} which is a consequence
of the chiral anomaly \citep{Adler:1969gk,Bell:1969ts}.
This can have a number of consequences.
There is an instability that can amplify a helical magnetic field \cite{JS97}.
It is now often referred to as the chiral plasma instability (CPI)
\cite{Akamatsu:2013pjd} and it causes the chiral chemical potential carrying
the chirality of the fermions to decay such that the total chirality
remains unchanged \citep{Kamada:2018tcs,Domcke:2022uue, 2022arXiv221112517C}.
Conversely, if a helical magnetic field decays, the chiral chemical
potential can increase \citep{Hirono+15,Schober+2020}.
Finally, when the chiral chemical potential balances the magnetic helicity
to produce vanishing total chirality of the system, which is realized in,
e.g., cosmological MHD after axion inflation \citep{domcke2018, Domcke+19,
Domcke:2022kfs}, the magnetic field can only decay.
It has been thought that the decay is triggered by the CPI
and that it would be therefore exponential \citep{domcke2018,Domcke+19}.
In this Letter, however, we show that this decay occurs only
in a power-law fashion.
This has consequences for explaining the baryon asymmetry of the Universe
\cite{Giovannini:1997eg,Giovannini:1997gp,Kamada:2016cnb} and for theories
of primordial magnetic fields, which will open up a new direction for
early Universe cosmology model building.
The purpose here is to show that the decay of the magnetic field in
chiral MHD is governed---similarly to nonhelical MHD---by a
conserved quantity that we call the adapted Hosking integral.
While the model adopted here is based on quantum 
electrodynamics, the extension to the realistic 
cosmological models based on the standard model of 
particle physics is straightforward;
see, e.g., Refs.~\cite{Kamada:2022nyt,Domcke:2022uue}.

The Hosking integral $I_{\rm H}$ is defined as the asymptotic limit
of the relevant magnetic helicity density correlation integral,
${\cal I}_{\rm H}(R)$, for scales $R$ which are large compared to the correlation
length of the turbulence, $\xiM$, but small compared to the system
size $L$.
The function ${\cal I}_{\rm H}(R)$ is given by
\begin{equation}
{\cal I}_{\rm H}(R)=\int_{V_R}\aver{h(\xx)h(\xx+\rr)}\, \dd^3r,
\end{equation}
where $V_R$ is the volume of a ball of radius $R$ and, in MHD,
$h=\AAA\cdot\BB$ is the magnetic helicity density with $\AAA$ being
the magnetic vector potential, so $\BB=\nab\times\AAA$.
Here, angle brackets denote averages over the volume $L^3$.

For relativistic chiral plasmas, on the other hand, we now amend the
magnetic helicity density with a contribution from the chiral chemical
potential $\mu_5$.
We work here with the scaled chiral chemical potential
$\mu_5\to\mu_5'=(4\alpha/\hbar c)\,\mu_5$, where $\alpha$ is the fine structure
constant, $\hbar$ is the reduced Planck constant, and $c$ is the speed of light.
Our rescaled $\mu_5'$ has the dimension of a wave number.
From now on, we drop the prime and only work with the rescaled chiral
chemical potential.
We also define the quantity $\lambda=3\hbar c\,(8\alpha/\kB T)^2$, where
$\kB$ is the Boltzmann constant and $T$ is the temperature.
We define the total helicity density
$h_{\rm tot}\equiv\AAA\cdot\BB+2\mu_5/\lambda$ and replace
$h\to h_{\rm tot}$ when defining the adapted Hosking integral.

Similarly to earlier studies of non-relativistic chiral plasmas ($\mu_5\to0$) with a helical magnetic
field, the case of a finite net chirality, $\aver{h_{\rm tot}}\neq0$,
is governed by the conservation law for $\aver{h_{\rm tot}}$.
Of course, when $\aver{h_{\rm tot}}=0$, it is still conserved, but it
can then no longer determine the dynamics of the system.
This is when we expect, instead, $I_{\rm H}$ to control the dynamics of
the decay.
As before, we define $I_{\rm H}={\cal I}_{\rm H}(R_*)$ for values of
$R_*$ for which ${\cal I}_{\rm H}(R)$ shows a plateau.
In the following, we focus on this case using numerical simulations
to compute the decay properties of a turbulent magnetic field and
the conservation properties of $I_{\rm H}$ using the total helicity in a
relativistic plasma.

Now setting $c=1$, the evolution
equations for $\AAA$ and $\mu_5$ are \citep{Roga_etal17}
\begin{eqnarray}
\frac{\partial\AAA}{\partial t}&=&\eta(\mu_5\BB-\JJ)+\uu\times\BB,
\quad\JJ=\nab\times\BB,
\label{dAAdt}
\\
\frac{\partial\mu_5}{\partial t}&=&-\frac{2}{\lambda}\eta(\mu_5\BB-\JJ)\cdot\BB
-\nab\cdot(\mu_5\uu)+D_5\nabla^2\mu_5,\;\;
\end{eqnarray}
where $\eta$ is the magnetic diffusivity, $D_5$ is the diffusion
coefficient of $\mu_5$, spin flipping is neglected here, and $\uu$ is the
velocity, which is governed by the compressible Navier-Stokes equations
\citep{Roga_etal17, Bran_etal17, BHKRS21}
\begin{eqnarray}
\frac{\DD\uu}{\DD t}&=&\frac{2}{\rho}\nab\cdot\left(\rho\nu\SSSS\right)
-\frac{1}{4}\nab\ln\rho+\frac{\uu}{3}\left(\nab\cdot\uu
+\uu\cdot\nab\ln\rho\right)
\nonumber \\
&-&\frac{\uu}{\rho}\left[\uu\cdot(\JJ\times\BB)+\eta \JJ^2\right]
+\frac{3}{4\rho}\JJ\times\BB,
\label{dudt} \\
\frac{\partial\ln\rho}{\partial t}
&=&-\frac{4}{3}\left(\nab\cdot\uu+\uu\cdot\nab\ln\rho\right)
+\frac{1}{\rho}\left[\uu\cdot(\JJ\times\BB)+\eta \JJ^2\right],
\nonumber\\
\label{dlnrhodt}
\end{eqnarray}
where ${\sf S}_{ij}=(\partial_i u_j+\partial_j u_i)/2
-\delta_{ij}\nab\cdot\uu/3$ are the components of the rate-of-strain
tensor, $\nu$ is the kinematic viscosity, $\rho$ is the density (which
includes the rest mass density), and the ultrarelativistic equation of
state for the pressure $p=\rho/3$ has been employed.
We assume uniform $\nu$, $\eta$, and $D_5$ such that $\nu=\eta=D_5$.
Our use of \Eqs{dudt}{dlnrhodt} compared to the nonrelativistic counterpart only
affects the kinetic energy and not the magnetic field evolution; see
Ref.~\cite{Bran+17} for comparisons in another context.

We define spectra of a quantity $h(\xx)$ as
$\Sp(h)=\oint_{4\pi}|\tilde{h}|^2\,k^2\dd\Omega_k/(2\pi L)^3$, where a
tilde denotes the quantity in Fourier space and $\Omega_k$ is the solid
angle in Fourier space, so that $\int\Sp(h)\,\dd k=\aver{h^2}$.
Here, $k\equiv|\kk|$.
The magnetic energy spectrum is $\EM(k,t)\equiv\Sp(\BB)/2$ and
$\int\EM\,\dd k=\aver{\BB^2}/2$ is the mean magnetic energy density.
The mean magnetic helicity density is $\HHM=\aver{\AAA\cdot\BB}$,
the magnetic helicity spectrum is $\HM(k,t)$ with $\int\HM\,\dd k=\HHM$,
and $\xiM=\EEM^{-1}\int k^{-1}\EM\,\dd k$ is the correlation length.

For an initially uniform $\mu_5\equiv\mu_{50}$, \Eq{dAAdt}
has exponentially growing solutions proportional to
$e^{\ii\kk\cdot\xx+\gamma_5 t}$, when $k<\mu_{50}$.
The maximum growth rate is $\gamma_5=\mu_{50}^2\eta/4$ for
$k=k_5\equiv\mu_{50}/2$ \citep{Roga_etal17, Bran_etal17}.
As an initial condition for $\AAA$, we consider a Gaussian distributed
random field with a magnetic energy spectrum
that is a broken power law
with $\EM(k,t)\propto k^4$ for $k<k_0$, motivated by causality constraints
\cite{DC03}, and a Kolmogorov-type spectrum, $\EM(k,t)\propto k^{-5/3}$,
for $k>k_0$, which may be expected if there is a turbulent forward cascade.
By setting $k_0=1$ for the spectral peak, we fix the units of velocity
and length.
The unit of time is then $(k_0)^{-1}$.
We initially set $\rho=\rho_0=1$, which then also fixes the units
of energy.

We solve the governing equations using the {\sc Pencil Code} \citep{JOSS},
where the equations are already implemented \citep{Schober+18, Schober+20}.
We consider a cubic domain of size $L^3$, so the smallest wave number
is $k_1=2\pi/L$.
The largest wave number is $k_{\rm Ny}=k_1 N/2$, where $N$ is the number
of mesh points in one direction.
In choosing our parameters, it is important to observe that
$k_1\ll k_0\ll k_5\ll k_{\rm Ny}$.
Here, we choose $k_1=0.02$, $k_0=1$, $k_5=5$, and $k_{\rm Ny}=10.24$,
using $N=1024$ mesh points in each of the three directions.
This means that $|\mu_{50}|=10$, which is virtually the same as
$k_{\rm Ny}$.
However, experiments with other choices, keeping $N=1024$, showed that
ours yields an acceptable compromise that still allows us to keep
$k_1$ small enough.
We choose the sign of $\mu_5$ to be negative, and adjust the amplitude
of the magnetic field such that $2\EEM\xiM=\HHM=-2\mu_{50}/\lambda$.
Using $\eta=2\times10^{-4}$ and $\lambda=2\times10^4$, we have,
following Ref.~\citep{Bran+17},
$v_\lambda\equiv\mu/\sqrt{\rho_0\lambda}\approx0.07$ and
$v_\mu\equiv\mu\eta=0.002$, so $v_\lambda/v_\mu\approx35\gg1$,
corresponding to what is called regime~I.

\begin{figure}[t!]\begin{center}
\includegraphics[width=\columnwidth]{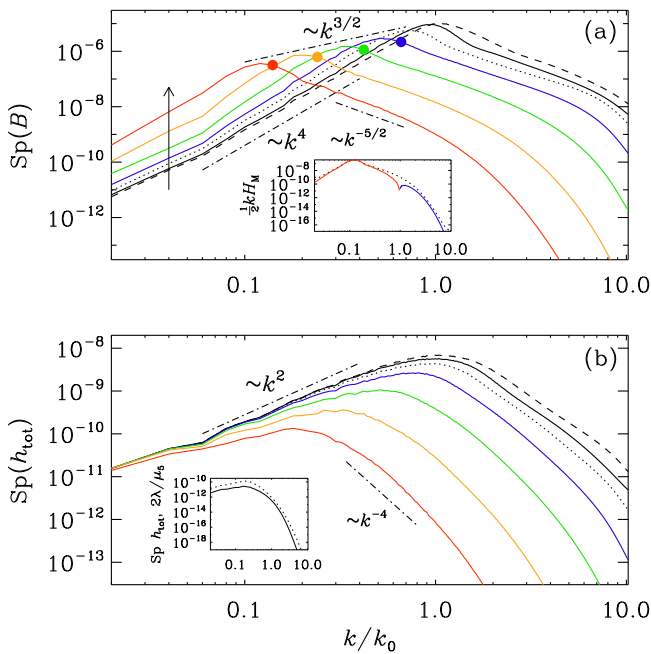}
\end{center}\caption[]{
(a) Magnetic energy and (b) total helicity variance spectra at $t=31$ (dashed),
$100$ (solid), $316$ (dotted), $10^3$ (blue), $3.16\times10^3$ (green),
$10^4$ (orange), and $3.16\times10^4$ (red).
In (a), note that $\Sp(\BB)$ evolves underneath the envelope $k^{3/2}$,
and the upward arrow indicates the sense of time.
For orientation, the slopes $k^{-5/2}$ and $k^{-4}$ have been indicated
in what is expected to correspond to the inertial ranges in (a)
and (b), respectively.
In (a), the inset shows $(k/2)\,\HM(k)$ at the last time with positive
(negative) values in red (blue), and in (b), the inset compares 
$\Sp(2\mu_5/\lambda)$ (solid) with $\Sp(h_{\rm tot})$ (dotted)
at the last time.
}\label{rspec_select_HoskM_1024a_mu10_k002b}\end{figure}

In \Fig{rspec_select_HoskM_1024a_mu10_k002b}(a), we present magnetic
energy spectra at different times.
We clearly see an inverse cascade where the spectral magnetic energy
increases with time for $k\ll k_0$ (indicated by the upward arrow),
but decays for $k\gg k_0$.
As time goes on, the peak of the spectrum moves to smaller wave numbers
with $k_{\rm peak}\approx\xiM^{-1}$, where $\xiM$ increases approximately
like a power law, $\xiM\propto t^q$, while the energy density decreases,
also approximately like a power law with $\EEM\propto t^{-p}$.
The spectral peak always evolves underneath an envelope
$\propto k^{3/2}$, which implies that $\max[\EM(k,t)]=\xiM(t)^{-\beta}$
with $\beta=3/2$, indicated by the upper dashed-dotted line in
\Fig{rspec_select_HoskM_1024a_mu10_k002b}(a).

\begin{figure}[t!]\begin{center}
\includegraphics[width=\columnwidth]{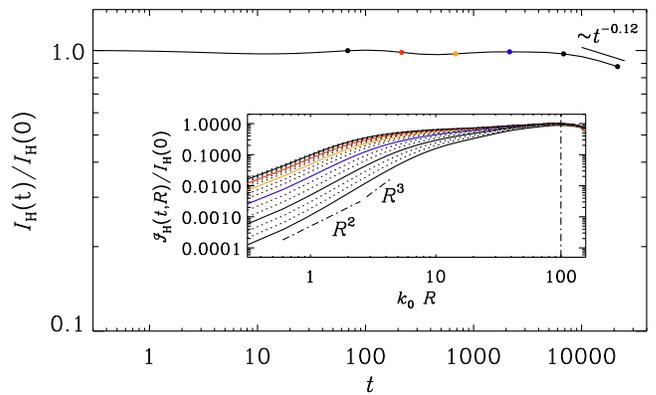}
\end{center}\caption[]{
$I_{\rm H}(t)$ normalized by its initial value.
The inset shows ${\cal I}_{\rm H}(R,t)$ versus $R$ at different times $t$:
solid lines correspond to $t=70$, 200, 700, 2000, 7000, and 20,000,
which are also marked by selected colored symbols in the graph of $I_{\rm H}(t)$.
The adapted Hosking integral is evaluated as $I_{\rm H}(t)={\cal I}_{\rm H}(R_*,t)$.
The vertical dashed-dotted line marks the value $k_0 R_*=100$ where the
curves show a plateau.
The slopes $\propto R^2$ and $\propto R^3$ are also marked by dashed-dotted lines.
}\label{psaff_1024a_mu10_k002b}\end{figure}

To compute ${\cal I}_{\rm H}$ (and thereby $I_{\rm H}$), we employ a
spectral technique by computing the total helicity variance spectrum
$\Sp(h_{\rm tot})$; see \Fig{rspec_select_HoskM_1024a_mu10_k002b}(b).
Compared to the inverse cascade seen in $\Sp(\BB)$, here we see the
conservation of the large-scale total helicity variance spectrum
$\propto k^2$.
We thus obtain
\begin{equation}
{\cal I}_H(R,t)=L^{-3}\int w(\kk,R)\,\Sp(h_{\rm tot})\,\dd^3 \kk/(2\pi)^3.
\label{weight}
\end{equation}
We choose $w(k,R)=(4\pi R^3/3)[6j_1(kR)/kR]^2$ as weight function
\citep{Zhou+22} with $j_n$ being spherical Bessel functions.

In \Fig{psaff_1024a_mu10_k002b}, we plot the adapted Hosking integral
$I_{\rm H}(t)$, normalized by its initial value.
It is evaluated as $I_{\rm H}(t)={\cal I}_{\rm H}(R_*,t)$ with
$k_0 R_*=100$, where ${\cal I}_{\rm H}(R,t)$ is shown in the inset
at different times as functions of $R$.
Note that $I_{\rm H}(t)$ is essentially flat and shows only toward the
end a slight decline $\propto t^{-0.12}$, which is similar to what
has been seen for other simulations at that resolution; see, e.g.,
Ref.~\cite{Bra23}.
Thus, the adapted Hosking integral appears to be well conserved --
even better so than the Hosking integral in ordinary MHD, studied in
Refs.~\citep{Hosking+Schekochihin21, Zhou+22}.
There is not even the slight uprise $I_{\rm H}(t)$ reported
in Ref.~\citep{Zhou+22}, which was there argued to be due to
strong non-Gaussian contributions to the field that emerged
during the nonlinear evolution of the system.
Note also that for $R\ll R_*$, we see
${\cal I}_{\rm H}(R,t)\propto R^2$, which is shallower
than the expected cubic scaling.
%JS: Should we indicate this scaling in the inset of Fig. 2?
%AB: added both now
%This is consistent with the work of Ref.~\citep{Zhou+22}, which
%showed an intermediate range $\propto R^2$, before cubic scaling
%emerged for even smaller $R$.
%AB: changed further
This might change at larger resolution, although an intermediate range
$\propto R^2$ is also seen in Fig.~4(d) of Ref.~\citep{Zhou+22}, before
cubic scaling emerged for $R/2\pi<10^{-4}$.

\begin{figure}[t!]\begin{center}
\includegraphics[width=\columnwidth]{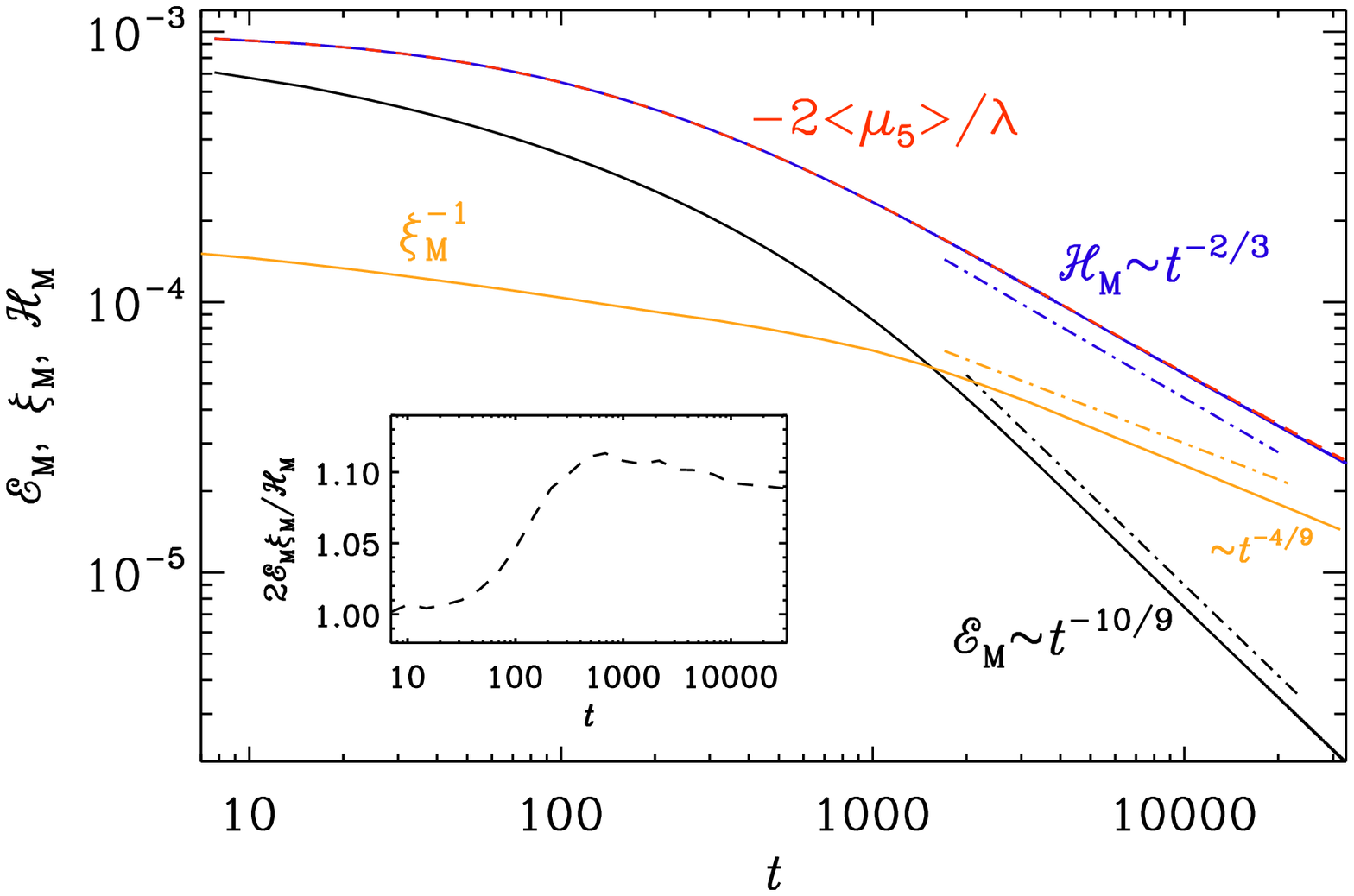}
\end{center}\caption[]{
Time dependence of $\EEM$ (black), $\xiM$ (orange), $\HHM$ (blue),
and $-2\aver{\mu_5}/\lambda$ (red).
The inset confirms that $2\EEM\xiM/\HHM\approx1$ during the whole time.
}\label{penerg_inset}\end{figure}

As in the case of nonrelativistic MHD ($\mu_5\to0$), the dimensions
of ${\cal I}_{\rm H}$ and $I_{\rm H}$ are $\cm^9\s^{-4}$.
This implies that in $\xiM\propto t^q$, the value of the exponent is
$q=4/9$, if the conservation of ${\cal I}_{\rm H}$ determines the time
evolution of the magnetic field around the characteristic scale.
Next, assuming selfsimilarity, the magnetic spectra can be collapsed on
top of each other by plotting them versus $k\xiM(t)$ and compensating the
decline in the height by $\xiM^\beta$ to yield the universal function
$\phi(k\xiM)=\xiM^\beta\EM(k\xiM)$; see Appendix~B of Ref.~\cite{Zhou+22}
and Refs.~\cite{BK17,Bra23} for examples in other contexts.
Using also the invariance of the spectrum under rescaling \citep{Ole97},
$\xx\to\xx'=\xx\ell$ and $t\to t'=t\ell^{1/q}$, and
since the dimension of $\EM(k,t)$ is $\cm^3\s^{-2}$, we have
$\EM(k\ell^{-1},t\ell^{1/q})=\ell^{3-2/q+\beta}[\xiM\ell]^{-\beta}\phi(k\xiM)$,
and therefore $\beta=2/q-3=3/2$, which agrees with
\Fig{rspec_select_HoskM_1024a_mu10_k002b}(a).
Finally, for $\EEM\propto t^{-p}$, we find with
$\EEM(t)=\int\EM\,\dd k\propto t^{-(\beta+1)q}$ the line $p=2(1-q)$,
which is also known as the self-similarity line \citep{BK17,Zhou+22}.
With $q=4/9$, we thus obtain $p=10/9$.
This is completely analogous to the MHD case with zero magnetic
helicity\footnote{See also the discussion in Ref.~\cite{Uchida:2022vue}
for weak magnetic field with zero magnetic helicity, where selfsimilarity
is not assumed.}; see also Table~2 of Ref.~\cite{Bra23}.
Thus, the cancellation of finite magnetic helicity by fermion chirality
with $\HHM(t)=-2\aver{\mu_5}(t)/\lambda\neq0$ has the same effect as that of
zero magnetic helicity.

To understand the decay of magnetic helicity density in the present
simulations, it is important to remember that the real space realizability
condition of magnetic helicity \citep{Kahn_etal13} is always valid and
implies $|\HHM|\leq2\EEM\xiM$.
Assuming the inequality to be saturated, we find the scaling
$|\HHM|\propto |\aver{\mu_5}|\propto t^{-r}$ with $r=p-q=2/3$.
This is well obeyed, as is shown in \Fig{penerg_inset}.
In the inset, we show that
$2\EEM\xiM/\HHM\approx1$ at early times and about 1.1 at late times.
It is thus fairly constant, therefore confirming the validity
of our underlying assumption.
On top of this evolution of the chiral asymmetry, the growth rate of the
CPI, $\gamma_5 \propto \aver{\mu_5}^2 \propto t^{-4/3}$, decays more rapidly
than $t^{-1}$, which causes it to grow less efficiently so as not to
spoil the scaling properties of the system.

\begin{figure}[t]\begin{center}
\includegraphics[width=\columnwidth]{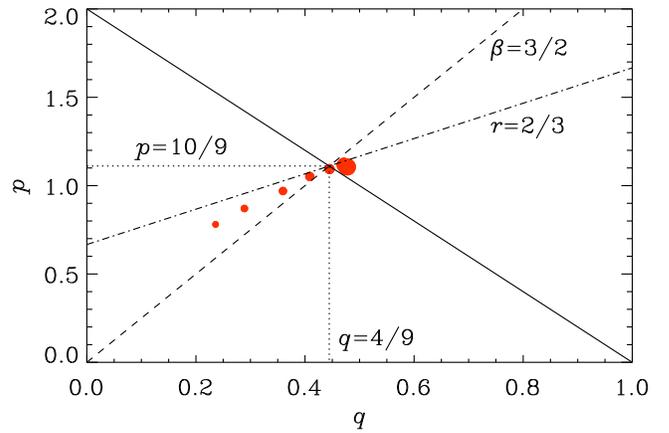}
\end{center}\caption[]{
$pq$ diagram for times $t=700$, 1000, 1500, 2200, 3200, 4600, 6800,
$10^4$, $1.5\times10^4$, $1.5\times10^4$, $2.2\times10^4$, and
$3.2\times10^4$, corresponding to symbols of increasing size.
The solid line denotes the scale-invariance line $p=2(1-q)$, the dashed
line the $\beta=3/2$ line for adapted Hosking scaling, and the dashed-dotted
line is the new $r=2/3$ line that does not have any correspondence in
standard MHD.
}\label{pEMxi_pq_runIb}\end{figure}

To characterize the scaling expected from the conservation of the adapted Hosking
integral further, in \Fig{pEMxi_pq_runIb} we plot the $pq$ diagram of
the instantaneous scaling exponents $p(t)=-\dd\ln\EEM/\dd\ln t$ versus
$q(t)=\dd\ln\xiM/\dd\ln t$.
The solution converges to a point close to the crossing point between
the $\beta=3/2$ line and the scale-invariance line $p=2(1-q)$.
The approach to the point $(p,q)=(10/9,\,4/9)$ does not occur
predominantly along the $\beta=3/2$ line, as in nonhelical standard MHD,
but is now closer to the $r=2/3$ line, where $p=q+r$.
In the unbalanced case, where the net chirality is non-vanishing,
however, the decay is solely governed by $\aver{h_{\rm tot}}=\const$
\cite{BKMSS23}.

In conclusion, we have presented evidence that, in the balanced case
of zero total chirality, the Hosking integral, when adapted to include
the chiral chemical potential, is approximately conserved around the
characteristic scale.
This implies decay properties for magnetic energy and correlation
length that are unchanged relative to nonhelical MHD, but here with
$\HHM+2\aver{\mu_5}/\lambda=0$ (instead of $\HHM=0$).
This yields the scaling $|\HHM|\propto|\aver{\mu_5}|\propto t^{-2/3}$,
along with the familiar scalings $\EEM\propto t^{-10/9}$ and $\xiM\propto t^{4/9}$ that
also apply to the case with $\HHM=0$.
These scalings have consequences for understanding the properties of
the chiral magnetic effect in the early Universe \citep{Kamada:2018tcs,
DelZanna+Bucciantini18, domcke2018,Domcke+19, Domcke:2022kfs} and young
neutron stars \citep{Masada+18,Dvornikov+20}.
Our work has significant impact on the baryon asymmetry of the Universe
from hypermagnetic helicity decay after axion inflation.
It also exposes a rather unexpected application of the general
idea behind the recently developed Hosking integral, raising therefore
the hope that there may be other ones yet to be discovered.

\medskip
\begin{acknowledgments}
We thank Valerie Domcke and Kai Schmitz for useful comments on the manuscript
and Kyohei Mukaida for fruitful discussions.
Support through Grant No.\ 2019-04234 from the Swedish Research Council (Vetenskapsr{\aa}det) (AB),
Grant-in-Aid for Scientific Research No.\ (C) JP19K03842 from the JSPS KAKENHI (KK), 
and Grant No.\ 185863 from the Swiss National Science Foundation (JS)
are gratefully acknowledged.
We acknowledge the allocation of computing resources provided by the
Swedish National Infrastructure for Computing (SNIC)
at the PDC Center for High Performance Computing Stockholm and Link\"oping.

\end{acknowledgments}

\bibliography{ref}{}
\end{document}